\documentclass{aa}
\nolinenumbers
\let\linenumbers\relax
\def\makeLineNumber{}
\usepackage{graphicx}
\usepackage{txfonts}
\usepackage{lipsum}

\usepackage{lscape}             % to rotate a single page table, example in appendix.
\usepackage{placeins}           % useful with \FloatBarrier, to keep 
\usepackage{txfonts}
\usepackage[]{hyperref}
\hypersetup{
    colorlinks=true,
    citecolor=blue,
    linkcolor=blue,
    urlcolor=blue,
    }
\newcommand{\kms}{km~s$^{-1}$}
\newcommand{\msun}{$M_{\odot}$}

\begin{document}
%%%%%%%%%%%%%%%%%%%%%%%%%%%%%%%%%%%%%%%%
% if you use custom commands in your title,
% ensure to check your title when submitting!
%%%%%%%%%%%%%%%%%%%%%%%%%%%%%%%%%%%%%%%%
\title{The nearby Type II supernova 2024ggi beyond 600 days: circumstellar interaction and dust formation}

%%%%%%%%%%%%%%%%%%%%%%%%%%%%%%%%%%%%%%%%
% Please separate each author with the \and command
%
% Use the \corrauth to provide the corresponding
% author address. It will be automatically inserted as 
% footnote in the PDF output.
%
% Please DO NOT include ORCIDs next to author names.
% Instead, please provide an active address for each coauthor:
% it will be automatically extracted by EDPS editorial system, 
% and co-authors will be be able to authenticate their ORCID.
%
% Only authenticated ORCIDs will be taken into account.
% ORCIDs included here will be removed.
%%%%%%%%%%%%%%%%%%%%%%%%%%%%%%%%%%%%%%%%

   \author{L. Ferrari\inst{1,2}\corrauth{luciaferrari@fcaglp.unlp.edu.ar}        % use \corrauth for the corresponding author
        \and G.~Folatelli \inst{1,2,3}
        \and C.~P.~Gutiérrez \inst{4,5}
        \and A.~Méndez \inst{1,2}
        \and C.~Ashall \inst{6}
        \and K.~Medler \inst{6}
        \and K.~Ertini \inst{1,2}
        \and H.~Kuncarayakti \inst{7,8}
        \and N.~Morrell \inst{9}
        \and M.~Stritzinger \inst{10}
        \and W.~Hoogendam \inst{6}
        \and E.~Hueichapán \inst{11,12}
        \and R.~Cartier \inst{13}
        \and J.~P.~Anderson \inst{14}
        \and M.~Bersten \inst{1,2,3}
        \and L.~Galbany \inst{4,5}
        \and J.~L.~Prieto \inst{12}
        \and C.~Cárdenas Uribe \inst{13}
        \and S.~Vazhayil \inst{12}
        }

    \institute{Instituto de Astrofísica de La Plata (IALP), CCT-CONICET-UNLP, Paseo del Bosque, B1900FWA La Plata, Argentina
    \and Facultad de Ciencias Astronómicas y Geofísicas, Universidad Nacional de La Plata, Paseo del Bosque S/N, 1900 La Plata, Buenos Aires, Argentina
    \and Kavli Institute for the Physics and Mathematics of the Universe (WPI), The University of Tokyo, 5-1-5 Kashiwanoha, Kashiwa, Chiba 277-8583, Japan 
    \and Institute of Space Sciences (ICE-CSIC), Campus UAB, Carrer de Can Magrans, s/n, E-08193 Barcelona, Spain.
    \and Institut d'Estudis Espacials de Catalunya (IEEC), 08860 Castelldefels (Barcelona), Spain
    \and Institute for Astronomy, University of Hawai`i at M\={a}noa,
2680 Woodlawn Drive, Honolulu, HI 96822, USA
    \and Finnish Centre for Astronomy with ESO (FINCA), 20014 University of Turku, Finland
    \and Tuorla Observatory, Department of Physics and Astronomy, 20014 University of Turku, Finland
    \and Las Campanas Observatory, Carnegie Observatories, Casilla 601, La Serena, Chile
    \and Department of Physics and Astronomy, Aarhus University, Ny Munkegade 120, DK-8000 Aarhus C, Denmark 
    \and Department of Astronomy, University of Florida, Bryant Space Science Center, Gainesville, FL 32611-2055, USA
    \and Instituto de Estudios Astrof\'isicos, Facultad de Ingenier\'ia y Ciencias, Universidad Diego Portales, Av. Ej\'ercito Libertador 441, Santiago, Chile
    \and Centro de Astronom\'ia (CITEVA), Universidad de Antofagasta, Avenida Angamos 601, Antofagasta, Chile
    \and European Southern Observatory, Alonso de Cordova 3107, Casilla
19, Santiago, Chile
    }
   % \date{Received October 7, 2026}

% \abstract{}{}{}{}{}
% 5 {} token are mandatory
 
  \abstract
   {The Type II supernova (SN) 2024ggi is one of the nearest core-collapse SNe to have exploded in recent decades. Its proximity has enabled extensive multiwavelength monitoring from early times through the nebular phase. Signatures of late-time interaction between the SN ejecta and hydrogen-rich circumstellar material (CSM) have recently been reported.
   Here we extend its study using late-time optical photometry and spectroscopy from 570 to 820~d after explosion.}
   {We investigate the emergence and evolution of late-time ejecta--CSM interaction and constrain the geometry of the interaction region and the properties of newly formed dust.} 
   {We present optical light curves and a series of low-resolution optical spectra of SN~2024ggi covering phases between 570 and 820~d after explosion.
   Comparisons with other interacting SNe II and with synthetic spectra are presented. We model the H$\alpha$ profile with the Monte Carlo radiative-transfer code \texttt{Damocles}, including both gas and dust components.}
   {Spectroscopic signatures of interaction precede the photometric flattening. The spectra evolve from ejecta-dominated to interaction-dominated as H$\alpha$ develops a broad, boxy profile. The optical light curves subsequently flatten relative to the $^{56}$Co decay rate at approximately 650--670~d. 
   The emission lines show a systematic blue-to-red asymmetry that requires a dust component newly formed within the ejecta.
   The inferred dust mass formed in the ejecta is $10^{-6}-10^{-5}~M_{\odot}$ if the grains are amorphous carbon or a carbon--silicate mixture, but can reach up to $10^{-3}~M_{\odot}$ for pure silicates.} %--cold dense shell (CDS) region
   {By $\gtrsim 600$~d, the optical evolution of SN~2024ggi is dominated by ejecta--CSM interaction, while the asymmetric line profiles provide evidence for recent dust formation.
   Our observations and findings of SN~2024ggi add to the growing number of SNe~II observed several hundred days after explosion, demonstrating the diversity in late-time interaction, and constraining the mass-loss histories of these massive star explosions.}


   \maketitle

%%%%%%%%%%%%%%%%%%%%%%%%%%%%%%%%%%%%%%%%%%%%%%%%%%%%%%%%%%%%%%

\section{Introduction}
\label{sec:intro}

Mass loss plays a critical role in determining the final evolution of massive stars and the observational properties of their explosions as supernovae (SNe). Type II SNe (SNe~II), the most common class of core-collapse SNe, are generally associated with red supergiant (RSG) progenitors that retain substantial H-rich envelopes at explosion \citep{Smartt2009,vandyk2012}.
However, canonical RSG winds cannot account for the
dense, confined circumstellar material (CSM) inferred around many SNe~II, material that must have been lost during the final years before explosion \citep[see, e.g.,][]{GonzalezGaitan2015,Forster2018}.

In early observations of SNe~II, interaction manifests as high luminosity, especially in the ultraviolet, and blue colours \citep{Dessart2022,Moriya2023,Jacobson-Galan2024,Ertini2026}, 
a delayed rise to maximum light \citep{Moriya2018,Forster2018}, 
radio and X-ray emission from non-thermal processes in the interaction region \citep{Chevalier2017,Matsuoka2019,Moriya2021}, high-velocity absorption features in the spectra \citep{Chugai2007,Gutierrez2017,Dessart2022}, 
and short-duration narrow emission lines from high ionization states, e.g. \ion{H}{i}, \ion{He}{ii}, \ion{C}{iv}, etc, in the optical spectra during the first days/week after the explosion \citep[also known as ``flash features'', ][]{Yaron2017,Bruch2023,Jacobson-Galan2024}. 
These observations probe CSM structures that typically extend over radii of $\sim10^{14}$--$10^{15}$~cm and, given typical RSG wind velocities, must have been formed during the last months to years before the explosion. Assuming steady-state winds, the implied mass-loss rates are $\dot{M}\sim10^{-3}$--$10^{-1}$~M$_\odot$~yr$^{-1}$, orders of magnitude higher than those measured in RSGs \citep{Beasor2020} and than predicted by standard prescriptions for single RSG winds \citep{deJager1988,vanLoon2005}.
Most massive stars are members of binary or multiple systems and are expected to interact with a companion during their evolution \citep{Sana2012}, making binary mass transfer and common-envelope episodes a plausible channel for producing such dense CSM. 
High mass loss rates have also been attributed to slowly accelerated winds \citep{Moriya2017} or to episodic pre-SN outbursts \citep[e.g.][]{Fuller2017}.

Ejecta--CSM interaction also leaves signatures at late times, most notably in the form of complex, flat-topped or boxy\footnote{That is, broad emission exhibiting sharp, roughly symmetric edges.} H$\alpha$ profiles, typically accompanied by a flattening of the light curve as the interaction adds to the declining radioactive luminosity \citep[e.g.][]{Dessart2023}. 
The epochs at which these features appear span a wide range. In several SNe~II with short plateaus, or with linear and fast-declining light curves, they emerge as early as $\sim$$200$--$500$~d after explosion \citep[see][]{Weil2020,Dessart2023}.
No such line profiles were reported between $\approx$150 and 500~d in the sample of 38 typical SNe~II studied by \citet{Silverman2017}. Among the objects followed to very late phases, only a handful of SNe~II with standard plateau durations have shown comparable signatures at epochs beyond $\approx$600~d. 
SNe become intrinsically faint at epochs of several hundred days and beyond, and observations at these phases therefore remain scarce. As a result, the fraction of SNe~II that undergo this transition from radioactive- to interaction-powered evolution has not yet been established.
Since mass loss is ubiquitous among massive stars, all core-collapse SNe should eventually interact with their CSM, although the interaction becomes observable only if the CSM is sufficiently dense. 

\begin{figure*}
    \centering
    \includegraphics[width=0.9\linewidth]{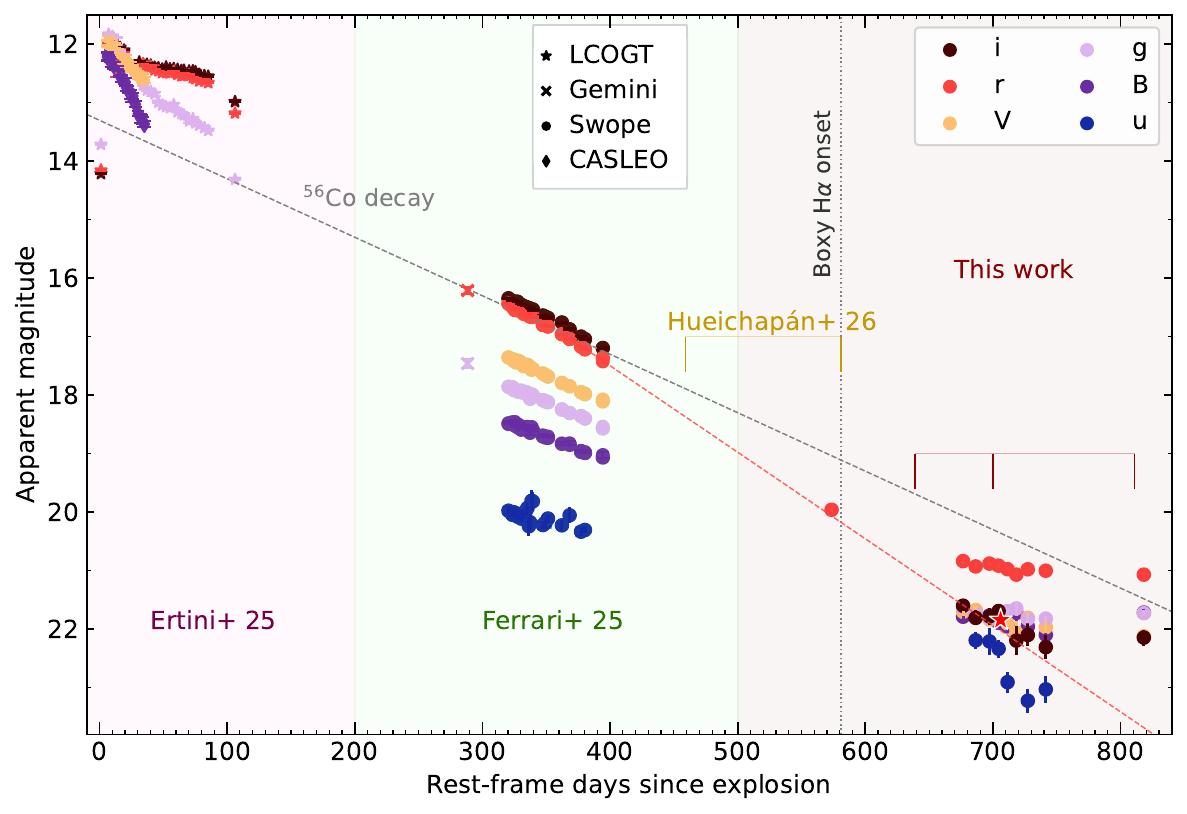}
    \caption{Observed light curves of SN~2024ggi from the explosion date until 820~d. Photometry earlier than 570~d was reported by \citet{Ertini2025} and \citet{Ferrari2025}. 
    The light curves flatten at $\sim650$~d, departing from the $^{56}$Co decay rate (grey line) and from the extrapolated $365-400$~d $r$-band decline (red line). 
    The red star at 706~d denotes the synthetic photometry of a subtracted spectrum where the H$\alpha$ line has been removed. 
    Red vertical lines mark the epochs of the spectra presented in this work and yellow vertical lines those of the two nebular spectra published by \citet{Hueichapan2026}. The first boxy H$\alpha$ profile is indicated with dotted gray line at 581~d, see discussion in Sect.~\ref{sec:spectra}.}
    \label{fig:LC}
\end{figure*}

Late-time interaction features probe CSM structures located at a further distance (than that where the material that creates the early narrow lines is found) of $\sim$$10^{15}-10^{16}$~cm from the progenitor star.
This material must have been expelled by the star during the final centuries to millenia before the explosion. 
The appearance of boxy H$\alpha$ profiles in the standard-plateau Type II SNe~2004et and 2017eaw has been associated with mass loss rates of $\dot{\mathrm{M}}\sim10^{-6}$~M$_\odot$~yr$^{-1}$ \citep{Kotak2009,Weil2020}, in agreement with estimates from radio and X-ray observations \citep{Misra2007,Szalai2019}.
This is well in line with the estimated mass loss rates for RSG stars.
Whether the confined CSM responsible for flash-ionisation features and the more distant CSM revealed by late-time interaction arise from a continuous, time-variable mass-loss history or from physically distinct processes remains an open question.

In the present work, we analyse the nebular spectral evolution of the nearby Type~II SN~2024ggi obtained between 630 and 811~d after the explosion. 
A boxy H$\alpha$ emission profile has fully developed, accompanied by a flattening of
the optical light curves. 
Such signatures were absent in the nebular spectra up to 400~d presented by \citet{Ferrari2025}, and had started to be present in the 581~d spectra published by \citet{Hueichapan2026}.
The emergence of these features between $\sim$460 and 580~d places SN~2024ggi among the small group of SNe~II with standard plateau durations that develop interaction signatures at ages earlier than $\approx$600~d. This is consistent with the relatively long plateau of $\approx$90~d measured for this SN \citep{Ertini2025}, and contrasts with the
case of SN~2023ixf, an equally nearby event that showed a shorter plateau of $\approx$80~d \citep{bersten2024} that developed the onset of a boxy H$\alpha$ profile as early as 309~d \citep{Medler2025_23ixf}.

Given its proximity, SN~2024ggi together with SN~2023ixf are two of the closest core-collapse SNe of the decade, and have been targets of extensive observational campaigns that constrained the nature of their progenitors and the properties of their CSM.
The progenitor of SN~2024ggi has been directly identified in pre-explosion Hubble Space Telescope \textit{(HST)} and \textit{Spitzer} imaging as an RSG with an initial mass of $\approx$13--15~M$_\odot$ \citep{Xiang2024_ggi,Hong2024}.
Early flash-spectroscopy observations revealed narrow, high-ionization emission lines arising from a dense, confined CSM extending to $\lesssim$$5\times10^{14}$~cm, associated with enhanced mass-loss rates of $\dot{\mathrm{M}}\sim10^{-3}-10^{-2}$~M$_\odot$~yr$^{-1}$ during the last years prior to the explosion
\citep{Jacobson-Galan2024,Zhang2024,Shrestha2024,Pessi2024}. 
Similar values were favoured by hydrodynamical modelling of the early-time light curve. These models point to a two-component CSM and a mass-loss rate of $\dot{\mathrm{M}}\sim4\times10^{-3}$~M$_\odot$~yr$^{-1}$ confined within $\sim3000$~R$_\odot$ \citep[$\sim2\times10^{14}$~cm;][]{Ertini2025,Chen2025}.
In contrast, lower rates of $\dot{\mathrm{M}}\sim6\times10^{-5}$~M$_\odot$~yr$^{-1}$ were inferred from the X-ray emission \citep{Ferdinand2026}, while the quiescent pre-explosion mass-loss rate derived from the progenitor spectral energy distribution (SED) is $\lesssim$$3\times10^{-6}$~M$_\odot$~yr$^{-1}$
\citep{Xiang2024}, orders of magnitude below the values required during the
final pre-SN years in order to explain the early flash-spectroscopy observations.

At larger distances from the progenitor, the CSM density is expected to drop, marking the transition from the enhanced mass loss of the final years to the more quiescent wind of earlier epochs. Late-time ultraviolet spectroscopy obtained at $\approx$230~d already revealed ongoing ejecta--CSM interaction, tracing material out to $\sim$$1.7\times10^{16}$~cm and a mass-loss rate declining to $\sim$$10^{-4}$~M$_\odot$~yr$^{-1}$ at look-back times of $\sim$150~yr \citep{Bostroem2026}.
The optical spectra presented here give an optical counterpart to that interaction, providing a complementary view on the outer CSM structure.
Infrared (IR) monitoring over comparable phases places tight constraints on when dust appeared.
The $0.4$--$21~\mu$m SED at $\approx$55~d shows no mid-IR excess \citep{Baron2025}, and NIRSpec spectroscopy at $+285$ and $+385$~d reveals abundant CO, declining from $\sim$$8.7\times10^{-3}$ to $\sim1.3\times10^{-3}$~M$_\odot$ as it cools from $\approx$2900 to $\approx$2500~K, while the symmetric hydrogen lines indicate no dust within the ejecta \citep{Mera2026}.
A growing SiO reservoir is nonetheless predicted to trigger condensation at later times, so that by $\sim$400~d the ejecta appear chemically pre-conditioned for, but not yet producing, dust.
The epochs analysed here lie well beyond that limit. They trace both the onset of dust formation and the extent of the CSM at even larger radii, and thus probe the progenitor’s mass-loss history over the final centuries to millennia.

\section{Observations}
The data presented in this paper comprise three low-resolution, optical spectra and $uBgVri$ light curves. The data set was compiled from several observing programmes described below.
Phases presented in this paper are referred to an explosion date of MJD $= 60410.8 \pm 0.3$, which corresponds to the midpoint between the first detection \citep[MJD = 60411.1, ][]{Tonry2024} and the last non-detection \citep[MJD = 60410.5, ][]{Killestein2024}. 
Observations were corrected by redshift adopting the heliocentric redshift of the host galaxy $\mathrm{z_{Hel}} = 0.0024$ \citep{Koribalski2004}, and by reddening adopting $E(B-~V)_{\mathrm{tot}} = 0.16$ mag, as derived by \citet{Pessi2024}. 
The distance to NGC~3621 derived by \citet{Paturel2002} is $D = 6.7 \pm 0.4$~Mpc, measured from the Cepheid period--luminosity relation.

The light curves were taken by the Precision Observations of Infant Supernova Explosions (POISE, \citealt{Burns2021}) collaboration during the 2026A campaign. The observations were carried out with the 1-m Henrietta Swope Telescope at Las Campanas Observatory, covering the $uBgVri$ bands between $574$ and $818$~d. The cadence is $7-13$~d over phases $676.5-741.4$~d. The images were processed with the POISE photometric pipeline, which conceptually follows the Carnegie Supernova Project (CSP) pipeline \citep{Contreras2010,Krisciunas2017}.
Host-subtracted photometry is presented in the natural system of the Swope Telescope. Light curves are displayed in Fig.~\ref{fig:LC} and detailed in Table \ref{tab:LC}.

The first epoch of our spectroscopic coverage corresponds to a phase of 639~d. The spectrum was obtained on 2026 January 10 (UT) with the Low Dispersion Survey Spectrograph 3 (LDSS-3, \citealt{Allington-Smith1994}) mounted on the $6.5$~m Magellan Clay Telescope at Las Campanas Observatory, Chile. The spectral resolution at $6300$ \AA~is $370$ \kms. The data were reduced using \texttt{IRAF}\footnote{\url{https://iraf.net/}} \citep{Cartier2026}.

Another spectrum was acquired by the $8-$m Gemini South Telescope with the Gemini Multi-Object Spectrograph (GMOS, \citealt{hook2004}) 706~d post-explosion, on 2026 March 17 (UT), through the Program GS-2026A-Q-309 (PI Ferrari). The choice of the R400 grating combined with the $1.5$ arcsec slit resulted in a spectral resolution of $500$ km s$^{-1}$ at $6000$ \AA. The data reduction was performed following standard procedures using \texttt{DRAGONS} \citep{Labrie2023}.

The final spectrum at 811~d was acquired with the Baade Telescope at Las Campanas Observatory on 2026 June 24 (UT), equipped with the Inamori-Magellan Areal Camera \& Spectrograph (IMACS, \citealt{Dressler2011}) and using the grism Gri-300-$17.5$. This configuration yielded a resolution of $250$ km s$^{-1}$ at 6000 \AA. The reduction and calibration were performed through standard procedures with \texttt{IRAF}.

\begin{figure}
    \centering
    \includegraphics[width=\linewidth]{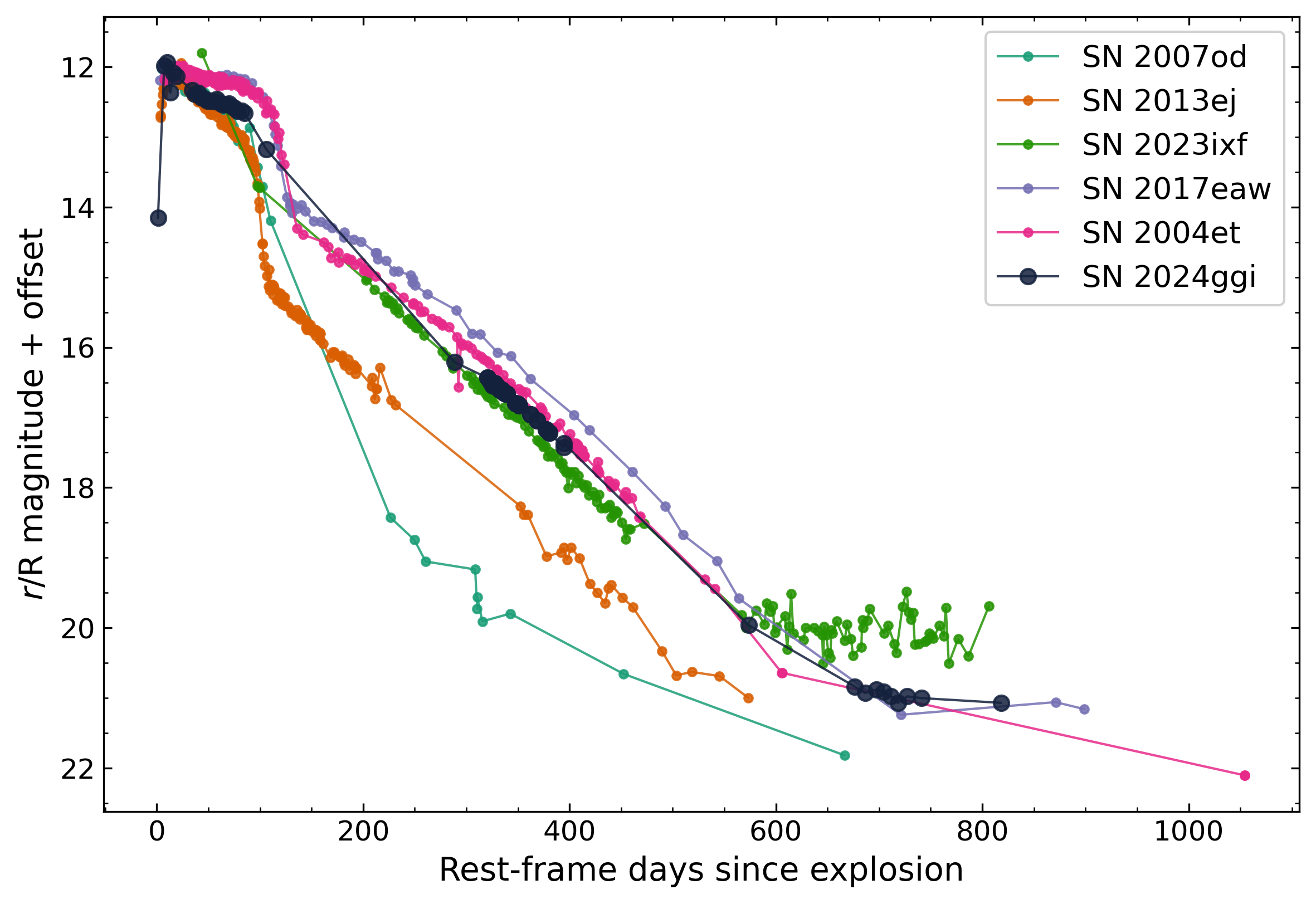}
    \caption{$r/R$-band photometry of late-time interacting SNe~II. All light curves were shifted to match the peak magnitude of SN~2024ggi. }%\textbf{23ixf to be included}
    \label{fig:LC_rband}
\end{figure}

\section{Analysis}

\subsection{Light curves}

The optical light curves of SN~2024ggi are shown in Fig.~\ref{fig:LC}. 
The late-time photometry reveals a flattening of the light curves, beginning
between 600 and 650~d.
This flattening is likely related to a transition from radioactive-decay power to interaction power.
We complement our data with the light curves published by \citet{Ertini2025} and \citet{Ferrari2025}, and indicate with vertical lines the epochs of the two spectra from \citet{Hueichapan2026} and of the three new spectra presented in this work, analysed in Sect.~\ref{sec:spectra}.

As discussed in \citet{Ferrari2025}, the light curves show a break at $\sim365$~d, when the decline rates steepen in the Swope $ri$ bands and in the Asteroid Terrestrial-impact Last Alert System (ATLAS; \citealt{Tonry2018, Smith2020}) $o$ band.
In Fig.~\ref{fig:LC} we also plot the $^{56}$Co~$\rightarrow$~$^{56}$Fe decay rate (gray line) and the extrapolation of the post-break $r$-band decline rate measured by \citet{Ferrari2025} (red line).
The single epoch at $574$~d photometry is consistent with the extrapolated decline, and beyond $\sim600$~d the observed evolution deviates from both, becoming essentially flat by $\sim650$~d after explosion. Another notable feature is the excess of the $r$ band over the remaining light curves, which reflects the dominance of H$\alpha$ in the spectrum (Sect.~\ref{sec:spectra}).

To quantify the contribution of H$\alpha$ to the $r$-band excess, the spectrum at $706$~d was flux-calibrated to  the Swope $r$-band photometry, adopting the CSP natural-system response function and zero point.
The H$\alpha$ emission within $\pm$8000~\kms\ of the rest wavelength was then subtracted, and the synthetic $r$-band magnitude was recomputed from the resulting line-free spectrum.
The line-free magnitude, $21.79$~mag (red star in Fig \ref{fig:LC}), is $0.88$~mag fainter than the observed photometry of the nearest epoch ($20.91$~mag at 704~d), implying that H$\alpha$ accounts for $\sim55$\% of the r-band flux at this epoch. 
As shown in Fig.~\ref{fig:LC}, this point falls close to the extrapolated post-break decline (red line), indicating that the underlying continuum flux fades at the pre-flattening rate and that the flattening seen beyond $\sim600$ is driven entirely by the interaction.

\begin{figure}[ht!]
    \centering
    \includegraphics[width=0.99\linewidth]{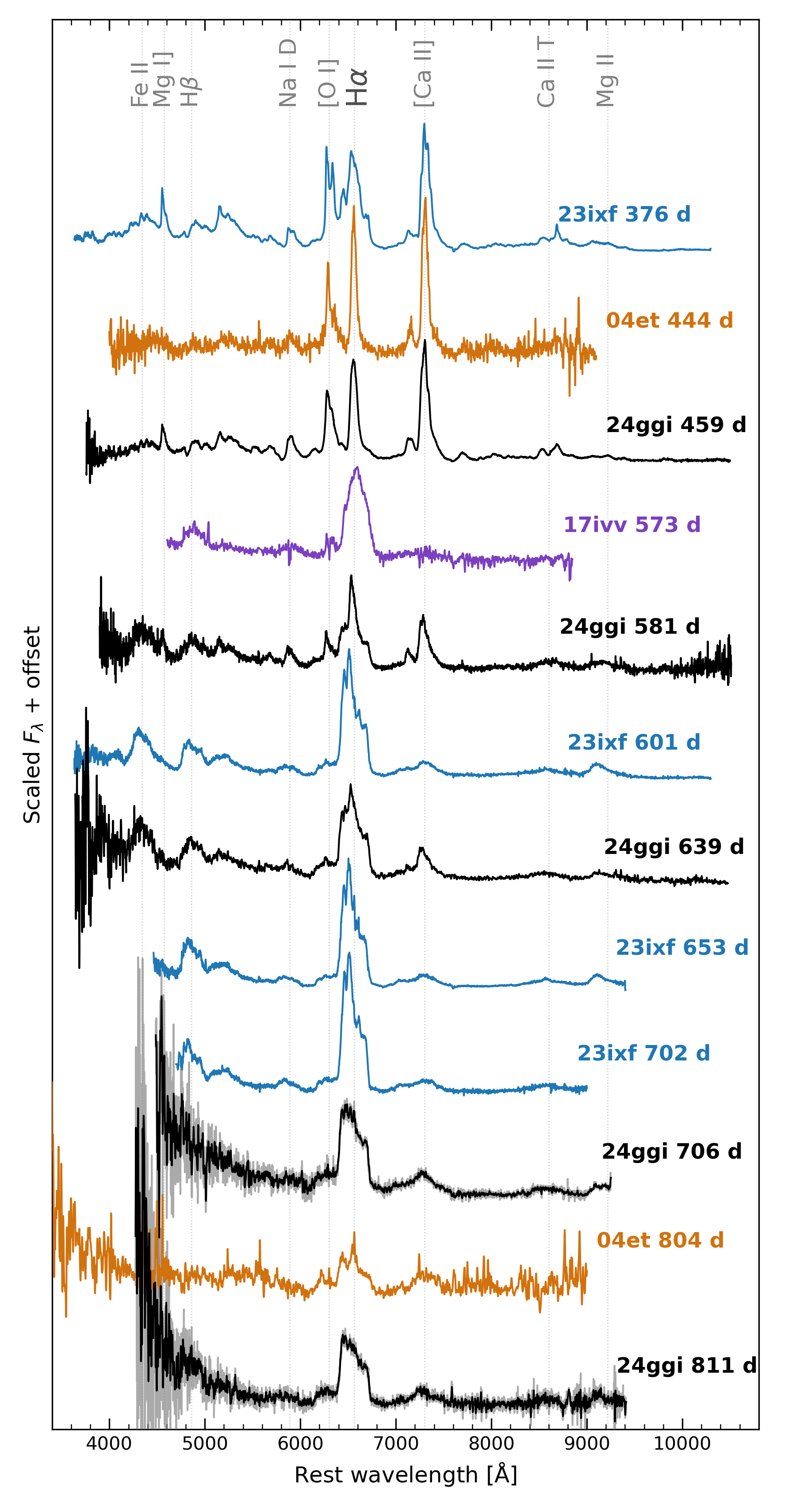}
    \caption{Nebular-phase spectral series of SN~2024ggi compared with other interacting SNe~II
    (SN~2004et, \citealt{Faran2014}; SN~2023ixf, \citealt{Folatelli2025} and \citealt{Medler2025_23ixf}). The 459 and 581~d spectra were published by \citet{Hueichapan2026}; the remaining three epochs are presented here for the first time. Spectra at 706 and 811~d have been smoothered for clearer visualization.}
    \label{fig:24ggi_spectralseries}
\end{figure}

Figure~\ref{fig:LC_rband} shows the $r/R$-band light curve of SN~2024ggi together with other SNe~II that developed interaction signatures at late times.
The temporal coverage does not allow precise epochs of the flattening to be determined, but the sequence appears to be led by SN~2007od \citep{Andrews2010}, followed by SN~2023ixf (downloaded from ATLAS\footnote{\url{https://fallingstar-data.com/forcedphot/}}, \citealt{Shingles2021}), and then by SN~2004et \citep{Misra2007,Maguire2010}, SN~2024ggi and SN~2017eaw \citep{Buta2019,Weil2020}.
In the case of SN~2013ej \citep{Bose2015,Yuan2016}, if a flattening occurs at all, it must take place beyond $\approx$600~d.
The spectroscopic evolution of SN~2007od and SN~2013ej nevertheless reveals the same boxy H$\alpha$ emission seen in the rest of the sample (Fig.~\ref{fig:24ggi_Ha_profiles}), which indicates that interaction is also present in these objects and that the shape of the late-time light curve alone is not a reliable diagnostic of its onset.
In SN~2004et, the absence of photometric observations between $\approx600$ and $\approx1100$~d prevents us from constraining the flattening epoch, although its
spectrum at $\approx840$~d already displays the characteristic boxy H$\alpha$ profile (\citealt{Faran2014}; see also Fig.~\ref{fig:24ggi_spectralseries}).

\subsection{Spectra}
\label{sec:spectra}

\begin{figure*}
    \centering
    \includegraphics[width=0.9\linewidth]{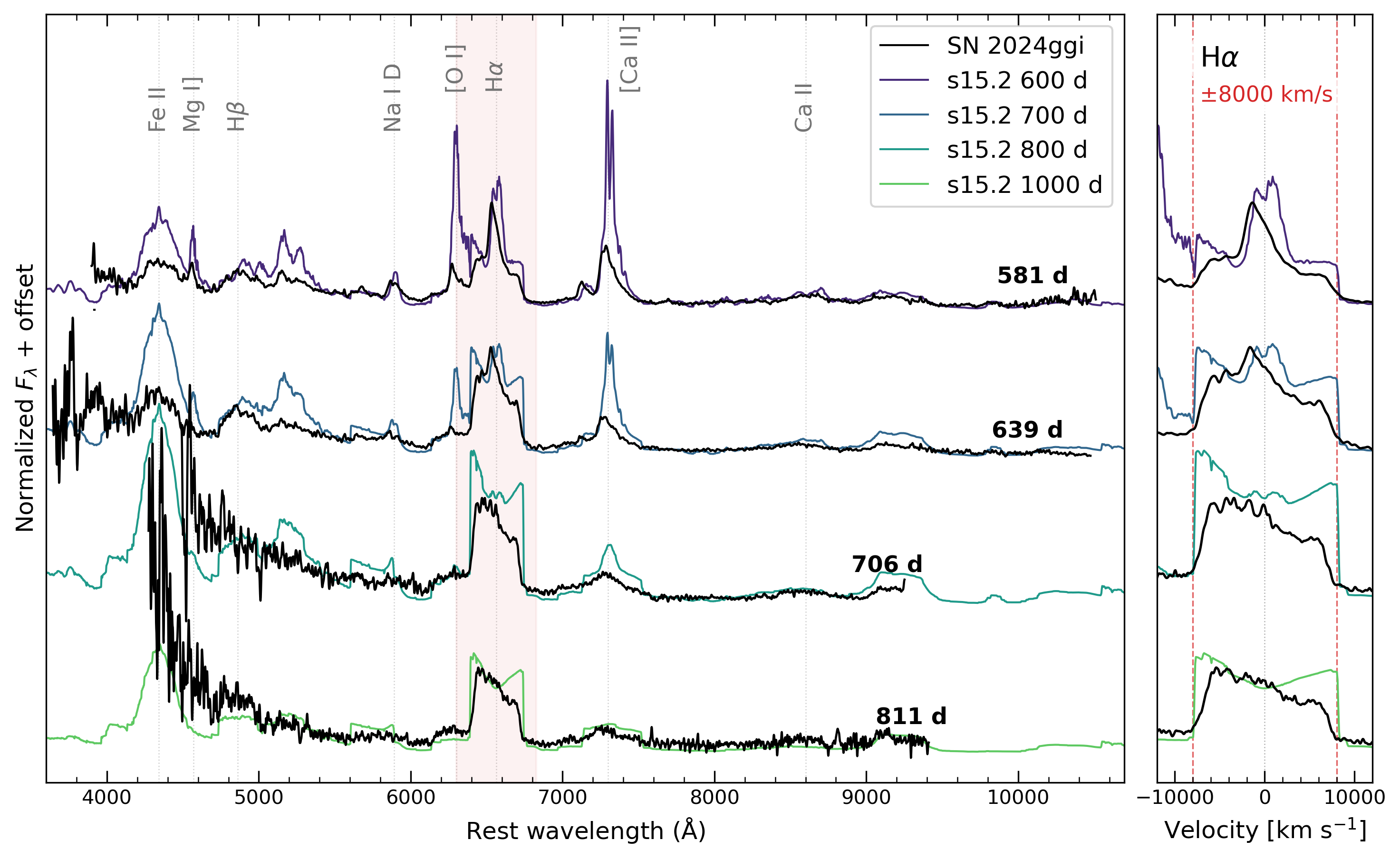}
    \caption{Left panel: Nebular spectra of SN~2024ggi at 581, 639, 706 and 811~d (black) compared with the
    interaction-powered SN~II models of \citet{Dessart2023} computed for shock powers of
    $1\times10^{40}$~erg~s$^{-1}$ (coloured lines, as labelled). Model phases do not necessarily correspond to the observed ones, since the shock power is injected at an arbitrary epoch.
    The pink region is zoomed in the right panel.
    Right panel: Zoom around the H$\alpha$ profile in velocity space. Vertical dashed lines mark the velocity at $\pm8000$ km s$^{-1}$.}
    \label{fig:24ggi_modelsDessart}
\end{figure*}

Fig.~\ref{fig:24ggi_spectralseries} shows the spectra of SN~2024ggi at 639, 706 and 811~d, together with those of other interacting SNe~II presented in \citet{Dessart2023}. 
The epochs at 459 and 581~d of SN~2024ggi published by \citet{Hueichapan2026} are included to trace the transition from an ejecta-dominated to an interaction-dominated regime.
As already noted by \citet{Hueichapan2026}, the defining feature of this transformation is the boxy H$\alpha$ emission. Present at every epoch from 581~d onwards, it progressively replaces the narrow, centrally peaked ejecta component, while most of the other nebular features fade.

The H$\alpha$ profile of SN~2024ggi closely resembles that of SN~2023ixf, except that the latter retains a central component ascribed to the unshocked inner ejecta rather than to the ejecta--CSM interaction. 
The flat-topped shape of H$\alpha$, with symmetric edge velocities, is the expected signature of emission from a geometrically confined, expanding shell.
We identify this shell with the cold dense shell (CDS) that forms between the forward and reverse shocks as the ejecta sweep up the CSM, and hereafter use ``CDS'' for the H$\alpha$-emitting interaction region and ``inner ejecta'' for the slower, unshocked material.
The two objects differ more markedly in the timing of the transition. 
In SN~2023ixf the boxy profile had already emerged by 309~d and was fully developed at 376~d \citep{Medler2025_23ixf}, whereas in SN~2024ggi it is still absent at 459~d and well established by 581~d. 
Interaction therefore came to dominate the optical spectrum roughly 200~d later in SN~2024ggi, which points to a more distant or less dense CSM structure. Between 581 and 811~d the signature continues to strengthen. 
The central peak from the inner ejecta weakens until it disappears, leaving a purely boxy profile whose red side is markedly depressed.
Since such an asymmetry requires dust located within or mixed with the emitting gas, this points to dust in the CDS itself (see Sect.~\ref{sec:damocles}). At 459 and 581~d, the blueshift is shared by the inner-ejecta lines ([O\,{\sc i}], [Mg\,{\sc i}], [Fe\,{\sc ii}]), which requires dust in the inner ejecta as well.
By 700~d the inner-ejecta emission has faded and the spectrum is dominated by the CDS, so the late-time asymmetry no longer constrains the dust in the inner ejecta.

The [\ion{O}{i}]~$\lambda\lambda6300,6364$ doublet evolves from the double-peaked profile seen at 581~d into a broader and weaker emission, again consistent with an interaction origin.
This additional contribution prevents us from estimating the progenitor mass at these phases, since the line flux no longer traces the oxygen-rich core alone.
H$\beta$ appears at a comparable epoch and with a velocity distribution similar to H$\alpha$. 
At 639~d we identify a broad emission blueward of \ion{Mg}{i}]~$\lambda4571$ (that has vanished by then) that is attributed to blended \ion{Fe}{ii} emission. 
At 706 and 811~d the low signal-to-noise ratio in the blue part of the spectra prevents us from establishing whether these features persist. 
From 581~d onward, broad emission develops around 9200~\AA, identified here as Mg~II $\lambda\lambda9218,9244$ and also reported in SN~2023ixf \citep{Folatelli2025}.
Its onset coincides with the emergence of the boxy H$\alpha$ profile and with the changes in the other lines, supporting an interaction origin for this feature as well.
The \ion{Ca}{ii}~IR has practically vanished at 639~d, whereas [\ion{Ca}{ii}]$\lambda\lambda7291,7324$ remains weak during all phases. %The forbidden doublet might be developing 

The spectral evolution of SN~2024ggi is compared with the late-time, interaction-powered SN~II models of \citet[][see Fig.~\ref{fig:24ggi_modelsDessart}]{Dessart2023}.
In those calculations, the ejecta--CSM interaction is mimicked by depositing energy in the form of non-thermal high-energy electrons within the CDS, placed at $v\approx8000$~km~s$^{-1}$, with shock powers in the range $1-7\times10^{40}$~erg~s$^{-1}$.
Two caveats must be kept in mind. 
First, the models include no dust, neither in the inner ejecta nor in the CDS, and therefore cannot reproduce line asymmetries or the associated flux deficit.
Second, the shock power is switched on at an arbitrary epoch ($350$~d), so model and observed phases need not correspond to one another. Only the sequence of spectral changes, not its absolute timing, can be meaningfully compared.

In the 581 and 639~d spectra, the \ion{Mg}{i}], [\ion{O}{i}], H$\alpha$ and [\ion{Ca}{ii}] emission all appear attenuated from the blue towards the red side of each line.
Since a red-wing suppression requires dust located within or mixed with the emitting gas, the asymmetry of lines formed in the slower inner ejecta ([\ion{O}{i}], \ion{Mg}{i}], [\ion{Ca}{ii}]) indicates newly formed dust in the inner ejecta. 
No such asymmetry is present at 459~d, which places the onset of dust formation in the inner ejecta between 459 and 581~d.
By 706 and 811~d the effect remains discernible only in H$\alpha$; the [\ion{O}{i}] doublet has by then adopted a boxy shape close to that of the interaction-powered models, albeit without their central peak.

\subsubsection{H\texorpdfstring{$\alpha$}~~profile}
\label{sec:subtraction}

In this section we analyse the H$\alpha$ profile at 706 and 811~d to constrain the properties of the emitting gas and the dust. We model the line with the Monte Carlo radiative-transfer code \texttt{Damocles} \citep{Bevan2016_damocles,Bevan2018_damocles}, designed to compute line profiles for a prescribed distribution of gas and dust.

\begin{figure}
    \centering
    \includegraphics[width=\columnwidth]{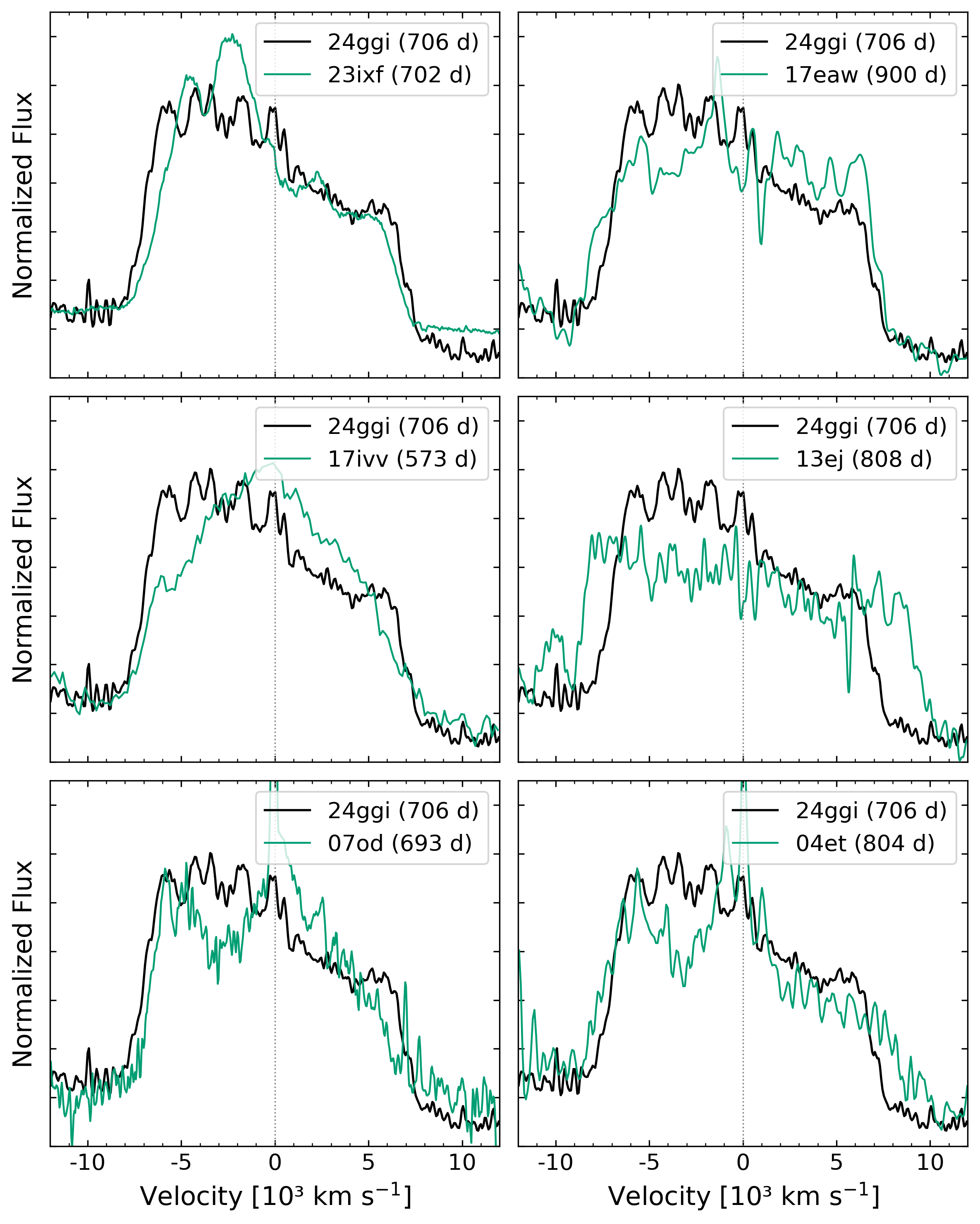}
    \caption{H$\alpha$ profile of SN~2024ggi at 706~d in velocity space, compared with those of other SNe~II showing late-time interaction signatures (SN~2023ixf, \citealt{Singh2026arXiv260314137S}; SN~2017eaw, \citealt{Weil2020}; SN~2017ivv, \citealt{Gutierrez2020}; SN~2013ej, \citealt{Mauerhan2017}; SN~2007od, \citealt{Andrews2010}; SN~2004et, \citealt{Faran2014}). The spectra were normalised to the overall flux of the boxy component. The epochs of SN~2024ggi and the comparison objects are indicated in the upper-right corner of each panel. Vertical line in each panel marks the 0 \kms~velocity.}
    \label{fig:24ggi_Ha_profiles}
\end{figure}

Fig.~\ref{fig:24ggi_Ha_profiles} compares the H$\alpha$ profile of SN~2024ggi with those of other SNe~II that developed interaction signatures at late times. 
The resemblance to SN~2023ixf and SN~2017eaw is remarkable, both in the flatness of the top and in the velocity extent of the boxy component.
SN~2017ivv reaches a comparable extent of $\approx\pm8000$~km~s$^{-1}$, but its profile is more rounded at the top, arising from both the radioactive heating of hydrogen-rich ejecta and the reverse shock interacting with high-velocity outer hydrogen-rich material \citep{Gutierrez2020}.
SN~2013ej is the only object in the sample with a significantly larger velocity, $\approx\pm10\,000$~km~s$^{-1}$, and also displays a flat top. 
SN~2007od and SN~2004et, in turn, show a composite profile, with a centrally peaked component (presumably formed in the inner ejecta) superimposed on the boxy emission.
In both cases, as in SN~2023ixf, the flux decreases steadily from the blue to the red side of the line, most likely because emission from the receding material is attenuated by a larger column of dust.
Taken together, this diversity of profile shapes and velocity extensions, and the wide range of epochs at which they become similar to one another, points to a correspondingly diverse set of CSM structures and distances from the progenitor. %shock front velocities?

\begin{figure}[ht!]
    \centering
    \includegraphics[width=.8\columnwidth]{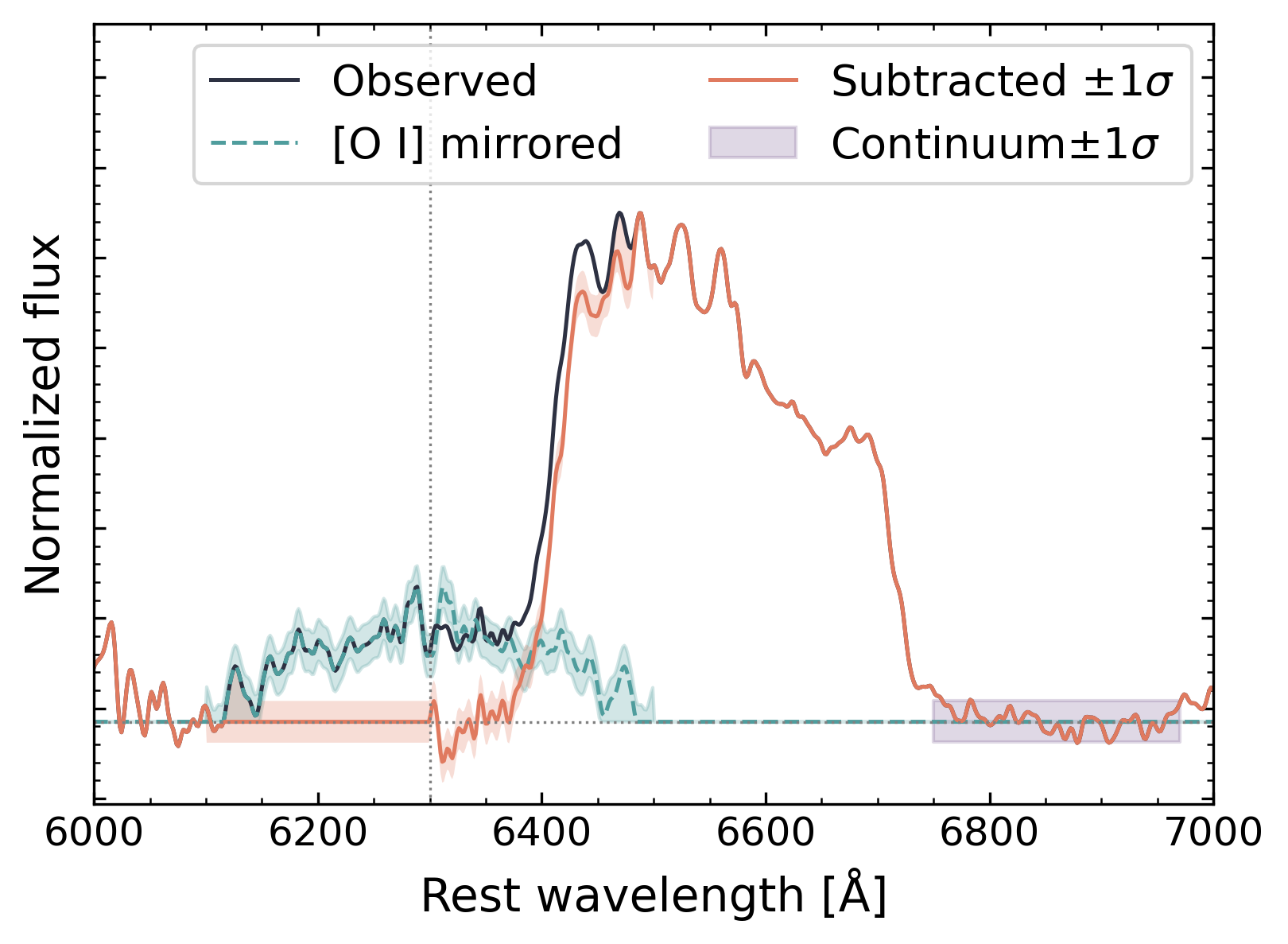}
    \caption{Scheme of the procedure used to clean the blue side of the H$\alpha$ emission from [\ion{O}{i}] doublet contamination. The black line is the smoothed spectrum at 706~d and the cyan line is the blue [\ion{O}{i}] profile mirrored to the red (dashed). The pink line shows the subtracted spectrum, assumed to be free of contamination. The shaded region indicates the continuum interval and its standard deviation; adopting the continuum $\pm1\sigma$ yields the shaded [\ion{O}{i}] profile and the corresponding range of subtracted spectra.}
    \label{fig:ggi_subtraction}
\end{figure}

Because of the large width of the [\ion{O}{i}] doublet, it contaminates the blue wing of H$\alpha$.
To isolate the H$\alpha$ emission, the [\ion{O}{i}]$\lambda\lambda6300,6364$ contribution is removed by mirroring its blue-side profile. This approach is preferable to an analytic fit because the profile is neither Gaussian nor strictly trapezoidal. A pseudo-continuum is then defined as the median of the flux in a region adjacent to H$\alpha$ that is free from other features, avoiding in particular [\ion{Fe}{ii}]$\lambda7155$ and [\ion{Ca}{ii}]$\lambda\lambda7291,7324$ emission.
The procedure is depicted in Fig. \ref{fig:ggi_subtraction}. 
The outcome is a spectrum that is flat immediately redward of 6300~\AA\ and shows a reduced flux between 6300~\AA\ and the inner part of the broad H$\alpha$ emission.
Contamination by [\ion{N}{ii}]~$\lambda\lambda6548,6583$ is not expected to affect this procedure: at the velocities involved, these lines fall within the central region of the boxy profile rather than near its edges, and no narrow residual is detected at their rest wavelengths.
The procedure is further justified by the models, where interaction-powered lines are symmetric with respect to their rest wavelength.

\begin{figure}[t]
    \centering
    \includegraphics[width=0.9\columnwidth]{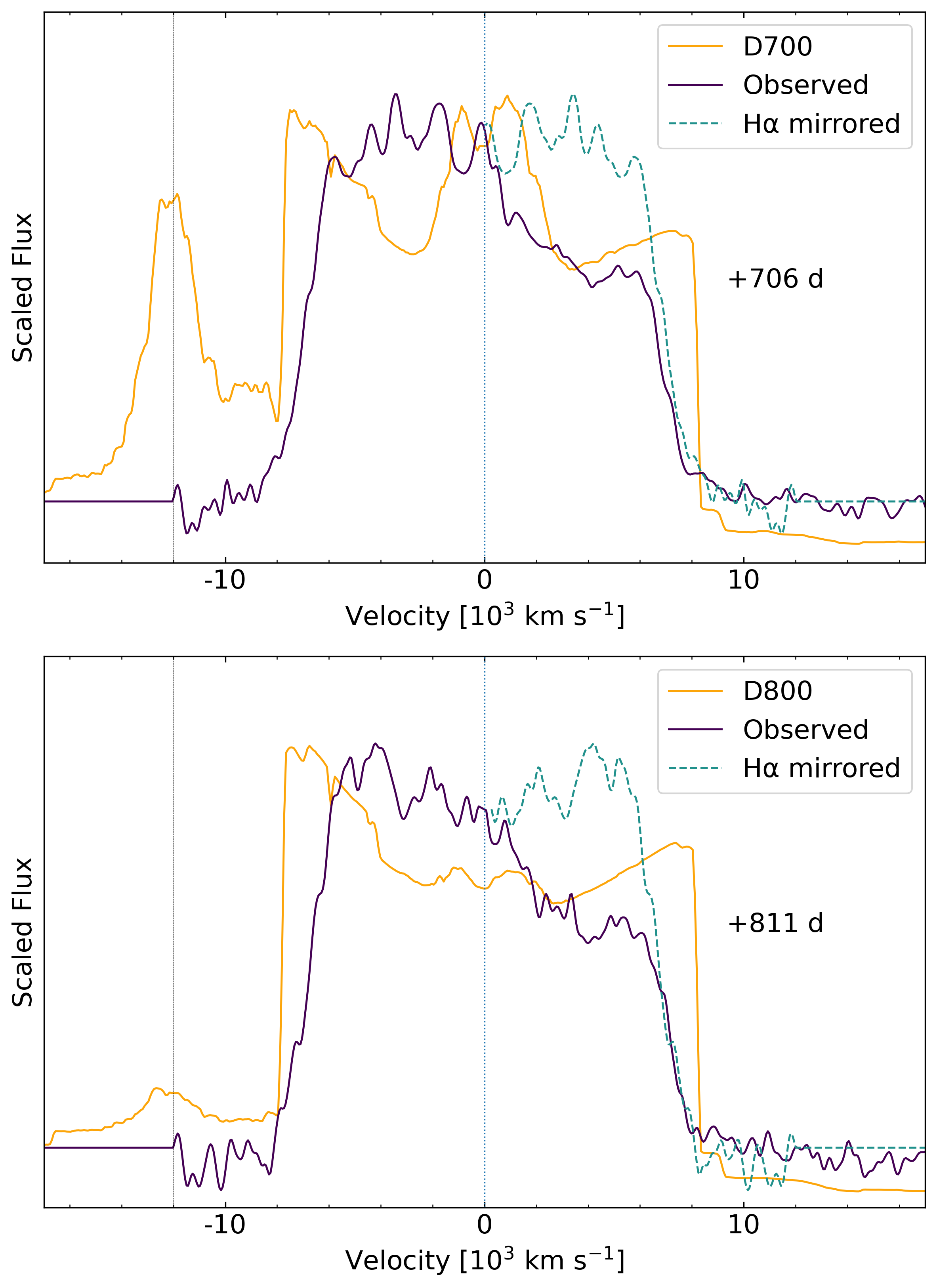}
    \caption{H$\alpha$ profile of SN~2024ggi at 706 (top) and 811~d (bottom), mirrored with respect to the rest
    wavelength of 6563~\AA. The models of \cite[][D700, top, D800, bottom]{Dessart2023} are shown for comparison. The blue and red edges of the observed profile match each other closely, and both are less steep than
    those of the models. Vertical line in each panel marks the 0 \kms~velocity.}
    \label{fig:ggi_mirror}
\end{figure}

Finally, the cleaned profiles at 706 and 811~d are mirrored with respect to the H$\alpha$ rest wavelength.
Figure~\ref{fig:ggi_mirror} shows both the smoothed and the subtracted spectra treated in this way.
The two edges of the boxy profile are found to be almost perfectly symmetric with respect to the rest wavelength, suggesting an overall spherical symmetry of the CDS.
For comparison, the figure also includes the interaction-powered models of \citet{Dessart2023} at the same nominal epochs, whose edges are appreciably steeper.
This difference is interpreted as evidence of a geometrically thicker CDS in SN~2024ggi than in the models.
A second distinguishing feature is the systematic decline of the flux from the blue to the red wing, again attributed to dust extinction and not reproduced by dust-free models.

The onset of the boxy H$\alpha$ profile at $\sim$581~d, for a shell expanding at 8000--10000~\kms, places the CSM swept up between 581 and 811~d at $\sim$(4--7)$\times10^{16}$~cm. 
For RSG wind speeds of 10--20~km~s$^{-1}$, this material was lost roughly $10^{3}$~yr before explosion. 
The UV shock power at $\sim$200~d \citep{Bostroem2026} shows that the ejecta were already interacting with CSM located at $\sim$$2\times10^{16}$~cm.
The optical signatures at 581~d therefore do not necessarily indicate the inner edge of a detached shell, and may instead reflect the epoch at which the interaction luminosity exceeds the declining radioactive input.

\subsubsection{Modelling with \texttt{Damocles}}\label{sec:damocles}

The H$\alpha$ line is modelled with the code \texttt{Damocles} \citep{Bevan2016_damocles,Bevan2018_damocles}, a Monte-Carlo radiative-transfer code developed to simulate the propagation of photons through a dusty, expanding ejecta of core-collapse SNe. 
Because photons emitted by the receding side of the emitting region pass through a larger dust column, dust preferentially extinguishes the redshifted emission, and thus the emerging line is blue-shifted.
The magnitude of this asymmetry is therefore a direct diagnosis of the dust mass, while the width and shape of the profile constrain the velocity and thickness of the CDS. 

The code is run over the parameter grid summarised in Table~\ref{tab:damocles}. The adopted values are motivated by the exploration and results presented in \citet{Niculescu-Duvaz2022}.
The dust species are allowed to vary between amorphous carbon dust, silicate dust, and a 50:50 mixture.
Grain sizes are set to $0.08-0.1$ $\mu$m.
The symmetric edges of the profile indicate a maximum velocity of $\sim$8000~\kms\ for the emitting region, and the explored range of $v_{\rm max}$ is chosen to bracket this value.
We do not assume homologous expansion, but rather let the outer radius $R_{\rm out}$ be independent of the maximum velocity.
The thickness of the CDS is denoted by the $R_{\rm in}/R_{\rm out}$ parameter, varied from $0.6$ to $0.9$.
The gas emissivity follows a radial power law, $j \propto r^{-\beta}$, where $\beta$ usually ranges from $1-3$ \citep{Niculescu-Duvaz2022}. 
In the input parameters, $\beta$ is named \texttt{gas\_emis}.
A wide range of dust masses is explored, from $10^{-6}$ to $10^{-3}$ \msun.
The dust is expected to be clumped, set by a low filling factor, taken as $0.1$ following \citet{Niculescu-Duvaz2022}. 
The dust and gas are treated as co-spatial and therefore share the same geometric parameters, namely the maximum velocity, outer radius, and shell thickness.
This assumption is motivated by the interpretation of the H$\alpha$ emission as arising in the CDS \citep{Pozzo2004,Bevan2020}, and by the fact that a dust shell external to the emitting gas would attenuate both sides of the line alike. 
Dust in the inner ejecta, inferred from the earlier spectra, is not included in these models. % (see below).

\begin{figure*}
    \centering
    \includegraphics[width=0.95\linewidth]{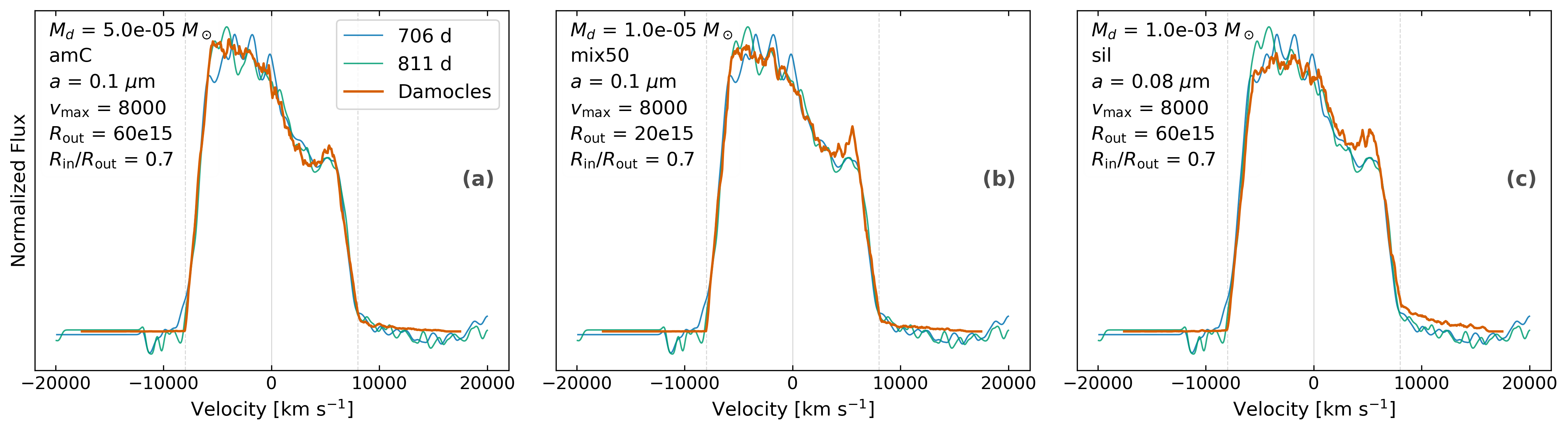}
    \caption{Best-fitting models that reproduce the H$\alpha$ line compared with the observed profiles at 706 and 811~d. Grey lines indicate the rest wavelength and $\pm 8000$\kms. Panel (a): best-fitting model among all parameters specified in Table \ref{tab:damocles}. 
    Panel (b): second best-fitting model, which happens to differ in the dust composition from pure amorphous carbonaceous to a 50:50 mixture with silicates.
    Panel (c): Best-fitting model among those with silicate dust only (see text for discussion).
    }
    \label{fig:24ggi_damocles}
\end{figure*}

As discussed in Sect.~\ref{sec:spectra}, the H$\alpha$ profiles at 706 and 811~d are quite similar. Thus, there is a set of parameters that match both profiles. The best-fitting models were selected by comparing them using a reduced $\chi^2$ statistic, computed over the velocity range $\pm10000$\kms around the rest wavelength of H$\alpha$ after the removal of the [\ion{O}{i}] contamination described in Sect. \ref{sec:subtraction}.
Each model is scaled to the observed profile by a multiplicative factor that minimizes the reduced $\chi^2$, so that the fit is sensitive to the shape of the line but not to its absolute flux.
Given the strong degeneracies among the model parameters, no single model is claimed to describe the exact distribution and properties of the material. Instead, the analysis distinguishes the quantities robustly constrained by the data from those
that remain unconstrained.

\begin{table}[ht!]
    \centering
    \caption{Summary of the \texttt{Damocles} input parameters and the values adopted to reproduce the H$\alpha$ line. Dust parameters marked with an asterisk (*) are tied to the corresponding gas parameter, since the two regions are assumed to be co-spatial. For parameters explored over a grid, the range limits and the number of equispaced points are given.}
    \label{tab:damocles}
    \begin{tabular}{l|c}
    Parameter    &  Range \\\hline
    \texttt{species} & \texttt{amC}; \texttt{mix50}; \texttt{sil} \\
    \texttt{grain\_a} (grain size) & $0.08;~0.1~\mu$m \\
    \texttt{gas\_vmax} ($v_{\rm max}$ & $7600-8800$ \kms~(4)  \\
    \texttt{gas\_rout} ($R_{\rm out}$)& $2-10\times10^{16}$ cm (3)\\
    \texttt{gas\_rratio} ($R_{\rm in}/R_{\rm out}$) & $0.6-0.9~(4)$ \\
    \texttt{gas\_emis}  ($\beta$ in $j \propto r^{-\beta}$) & $1.5-2.5~(3)$ \\
    \texttt{dust\_mass}  & $1.0\times10^{-6}-1.0\times10^{-3}$~\msun~$(7)$\\
    \texttt{dust\_ff} (filling factor) & $0.1$ \\
    \texttt{dust\_vmax*}  & $7600-8800$ \kms~(4) \\
    \texttt{dust\_rout*}  & $2-10\times10^{16}$ cm (3) \\
    \texttt{dust\_rratio*}  & $0.6-0.9~(4)$ \\
    \hline
    \end{tabular}
\end{table}

Figure \ref{fig:24ggi_damocles} shows the best-fitting models on top of the observed lines.
In all cases, the thickness of the shell follows the relation $R_{\rm in}/R_{\rm out}=0.7$.
This implies that the CDS extends through a relatively thick shell, needed to produce the trapezoid edges of the profile.
The best-fitting $R_{\rm out}$ is coincident with homologous expansion at $10000$ \kms, larger than the current maximum velocity of $\sim$8000~\kms. 
This indicates that the shell moved faster at earlier times and has since been decelerated by the CSM.

The dust mass required to reproduce the profile depends on the assumed composition.
If the dust were composed of amorphous carbon, the required mass would lie in the range $5\times10^{-6}$--$5\times10^{-5}$ \msun~(see Fig.~\ref{fig:24ggi_damocles}, panel a), with a 50:50 carbon--silicate mixture giving comparable values (panel b).
If it were instead composed of silicates, the mass would increase to $1.0\times10^{-3}$ \msun~(panel c).

\citet{Mera2026} report a decrease in the CO mass of SN 2024ggi between 285 and 385~d, which could originate from the dissociation of the molecules by the intense radiation field. In this scenario, the released carbon would be available to produce carbon dust, consistent with the carbon-rich model. 
On the other hand, \citet{Subrayan2026} recently modelled the mid-IR emission of SN~2024ggi at $660$~d, and found it to be consistent with a warm outer dust component (formed before the explosion) plus a cool, optically thin component composed of silicates newly formed in the ejecta.
Their derived mass of $\sim1.44\times10^{-3}$ \msun~is in good agreement with the value obtained here for a purely silicate composition, but exceeds by two orders of magnitude the mass required if the grains are carbonaceous or a $50$\% mixing of both.
Our modelling of the H$\alpha$ profile alone cannot discriminate between these compositions, since all reproduce the attenuation of the red wing.

The inferred dust mass, between $\sim$$10^{-5}$ and $\sim$$10^{-3}$~\msun\ depending
on composition, is typical of core-collapse SNe (CCSNe) at 700--800~d. In SN~1987A, line-profile modelling
yields $\sim$$3\times10^{-3}$~\msun\ at 615~d \citep{Bevan2016_damocles}. 
In SN~2023ixf, James Webb Space Telescope (JWST) observations revealed a cool silicate component of $\sim$$5.5\times10^{-4}$~\msun\ with no
significant growth between 250 and 720~d \citep{Medler2025_23ixf}, while models of its late-time spectra require $\sim$(0.5--1)$\times10^{-3}$~\msun\ of silicate dust in the CDS or inner ejecta \citep{JacobsonGalan2025_ixf}. 
{In SN~2004et, estimates at 650--800~d range from $\sim$$1.5\times10^{-4}$~\msun\
\citep{Kotak2009} to $\sim$$1.5\times10^{-3}$~\msun\ \citep{Fabbri2011}, depending on
the treatment of clumping. In the Type~IIn SN~2010jl, where dust also forms in the
post-shock region but the CSM is far denser, the dust mass had already reached
$\sim$$2\times10^{-3}$~\msun\ by $\sim$870~d \citep{Bevan2020}. 
These early masses fall two to three orders of magnitude short of the
$\sim$0.1--1~\msun\ per SN required for CCSNe to contribute appreciably to the dust
budget of high-redshift galaxies \citep{Dwek2007}.
Such masses are reached only after decades of continued grain growth, as seen in SN~1987A, where the dust mass reaches $\sim$0.8~\msun\ \citep{Matsuura2015,Bevan2016_damocles}. 
Line-profile modelling of a large CCSN sample indicates that dust formation proceeds on $\sim$30-yr time-scales and saturates at $\sim$0.4~\msun\ \citep{Niculescu-Duvaz2022}. The dust detected here is therefore consistent with the early stage of this growth, and SN~2024ggi offers the opportunity to follow it over the coming decades.

\section{Conclusion}
\label{sec:conclusions}

This work presents late-time optical photometry of SN~2024ggi up to 820~d after explosion, together with new nebular spectra at 639, 706 and 811~d. 
Combined with the two earlier epochs of \citet{Hueichapan2026}, these data trace the transition from radioactive- to interaction-powered emission.

From $\sim$650~d, the light curves depart from both the $^{56}$Co decay rate and the
decline measured after 365~d by \citet{Ferrari2025}, flattening in all bands.
The $r$ band develops an excess over the other bands as H$\alpha$ increasingly
dominates the spectrum. Removing H$\alpha$ from the flux-calibrated spectrum recovers
the pre-flattening $r$-band decline, showing that the continuum keeps fading while
the line does not.

The spectra evolve accordingly. A boxy H$\alpha$ component, absent at 459~d, emerges by 581~d and dominates all later epochs. Meanwhile, the central inner-ejecta component disappears and most other nebular features fade. 
The [\ion{O}{i}]~$\lambda\lambda6300,6364$ doublet turns from double-peaked into a broader,
weaker feature, thus no estimate of the progenitor mass can be derived from these spectra.
H$\beta$ emerges with a velocity distribution similar to that of H$\alpha$, and a broad \ion{Mg}{ii}~$\lambda\lambda9218,9244$ feature sets in. 
All these changes set in at roughly the same epoch, supporting a common interaction origin.

Interaction therefore began to shape the optical emission of SN~2024ggi some 150--200~d later than in SN~2023ixf, where the boxy H$\alpha$ profile had already emerged by 309~d \citep{Medler2025_23ixf}. This delay points to a more distant or less dense CSM structure.
% Dust is also clearly present. 
In the 581 and 639~d spectra, the \ion{Mg}{i}], [\ion{O}{i}], H$\alpha$ and [\ion{Ca}{ii}] show an attenuation of the red side of the emission profile.
Since it affects lines formed in the inner ejecta, this asymmetry indicates dust formed in the inner ejecta after $\sim$400~d.
By 706 and 811~d, when the inner ejecta emission has faded and the CDS dominates the spectrum, the effect remains discernible only in H$\alpha$, now arising from dust within the CDS.
The models of \citet{Dessart2023} reproduce the qualitative sequence of spectral changes but fail to account for two aspects of the observations. 
First, the edges of the observed boxy profile are appreciably less steep than in the models, pointing to a geometrically thicker emitting shell.
Second, the systematic blue-red flux asymmetry is absent from the models, computed without dust.

Once the [\ion{O}{i}] contamination of the blue wing is removed, the boxy H$\alpha$ emission extends to $\pm 8000$ km~s$^{-1}$ on both sides, consistent with a largely symmetric CDS.
Within the comparison sample, the shape of the late-time H$\alpha$ profile and its epoch of emergence vary markedly from object to object, ranging from the nearly flat profiles of SN~2013ej and SN~2023ixf to the composite profiles of SN~2007od and SN~2004et. 
This diversity reflects a correspondingly diverse set of circumstellar structures and distances from the progenitor.
In SN~2024ggi, the emergence of the boxy profile at $\sim$581~d places the CSM swept up between 581 and 811~d at $\sim$(4--7)$\times10^{16}$~cm, corresponding to material lost roughly $10^{3}$~yr before explosion.

The dust-induced blueshift of the H$\alpha$ profile at 706 and 811~d was modelled with \texttt{Damocles} over a grid of grain species, geometries, and dust masses.
The two epochs are reproduced by the same set of parameters, consistent with the similarity of their profiles.
Although the solutions are strongly degenerate, a few quantities are robustly constrained: the emitting region is a thick CDS with $R_{\rm in}/R_{\rm out}\simeq0.7$ expanding at $\sim8000$ \kms, as required by the trapezoidal edges of the line, and its outer radius agrees with material in homologous expansion at $10000$ \kms.
Amorphous carbonaceous grains, or a 50:50 carbon--silicate mixture, reproduce the profile with a dust mass of $5\times10^{-6}$--$5\times10^{-5}$ \msun, while silicate-only models require $\sim10^{-3}$ \msun.
This degeneracy is expected, since the line width is set by the kinematics of the CDS while the attenuation depends on the product of dust mass and optical opacity: the weaker optical absorption of silicates is compensated by a larger mass.
The H$\alpha$ profile alone therefore cannot discriminate between the two compositions, and independent evidence points in both directions: the decrease in CO mass between $285$ and $385$~d reported by \citet{Mera2026} is consistent with the condensation of carbon grains, whereas the silicate mass we obtain matches the cool dust component inferred from the mid-IR emission at 660~d by \citet{Subrayan2026}.
The dust mass inferred is comparable to that of other core-collapse SNe at similar epochs. It falls two to three orders of magnitude below the masses needed for CCSNe to contribute significantly to the dust budget of high-redshift galaxies, and is consistent with the early stage of a grain growth that proceeds over decades.

The proximity of SN~2024ggi makes it one of the few events that can be monitored for years or even decades. 
Continued spectroscopic and photometric follow-up, ideally extended to the ultraviolet and the mid-infrared, will constrain the outer extent of the CSM and the subsequent growth of the newly formed dust, providing further clues about the final stages of RSG evolution.

\begin{acknowledgements}
    L.F. acknowledges financial support from Consejo Nacional de Investigaciones Científicas y Técnicas (CONICET).
    C.P.G. acknowledges financial support from grant RYC2024-050959-I, funded by MICIU/AEI/10.13039/501100011033 and FSE+, and from grants PID2025-172154NA-I00 and PID2023-151307NB-I00, funded by MICIU/AEI/10.13039/501100011033, FSE+, ERDF, and the EU. Additional support was provided by CSIC through PIE 20215AT016, the María de Maeztu Unit of Excellence programme (CEX2020-001058-M), and the MaX-CSIC Excellence Award (MaX4-SOMMA-ICE).
    C.A. acknowledges support from NASA grants JWST-GO-03726, JWST-GO-05057, JWST-GO-05290, JWST-GO-06023, JWST-GO-06677,  JWST-GO-06583. Support for programs \#3726, \#5057, \#5290, \#6023, \#6677, and \#6583, were provided by NASA through a grant from the Space Telescope Science Institute, which is operated by the Association of Universities for Research in Astronomy, Inc., under NASA contract NAS 5-03127.
    M.D. Stritzinger is funded by the Independent Research Fund Denmark (IRFD, grant number  10.46540/2032-00022B). 
    W.B.H. acknowledges support from the National Science Foundation Graduate Research Fellowship Program under Grant No. 2236415. Any opinions, findings, conclusions, or recommendations expressed in this material are those of the authors and do not necessarily reflect the views of the National Science Foundation.
    E.H. was financially supported by Becas-ANID scholarship \#21222163.
    L.G. acknowledges financial support from MCIN and AEI 10.13039/501100011033 under project PID2023-151307NB-I00.
    This project has received funding from the European Union’s HORIZON-MSCA-2021-SE-01 Research and Innovation programme under the Marie Sklodowska-Curie grant agreement number 101086388.
    %Emilio
    This paper includes data gathered with the 6.5-meter Magellan Telescopes located at Las Campanas Observatory, Chile under the program allocated by the Chilean Telescope Allocation Committee (CNTAC), no: CN2025B-129 (P.I. E. Hueichapan).
    This work has made use of data from the Asteroid Terrestrial-impact Last Alert System (ATLAS) project. The Asteroid Terrestrial-impact Last Alert System (ATLAS) project is primarily funded to search for near earth asteroids through NASA grants NN12AR55G, 80NSSC18K0284, and 80NSSC18K1575; byproducts of the NEO search include images and catalogs from the survey area. This work was partially funded by Kepler/K2 grant J1944/80NSSC19K0112 and HST GO-15889, and STFC grants ST/T000198/1 and ST/S006109/1. The ATLAS science products have been made possible through the contributions of the University of Hawaii Institute for Astronomy, the Queen’s University Belfast, the Space Telescope Science Institute, the South African Astronomical Observatory, and The Millennium Institute of Astrophysics (MAS), Chile.
\end{acknowledgements}
% Please note that we have included the references below in
% order to compile the document, but we ask you to:

\bibliographystyle{aa_url} % style aa.bst
\bibliography{biblio} % your references Yourfile.bib

%%%%%%%%%%%%%%%%%%%%%%%%%%%%%%%%%%%%%%%%%%%%%%%%%%%%%%%%%%%%%%%
% Appendices must be placed after   \end{thebibliography}
% They will be placed automatically on a new page.
%%%%%%%%%%%%%%%%%%%%%%%%%%%%%%%%%%%%%%%%%%%%%%%%%%%%%%%%%%%%%%%
\begin{appendix}
%%%%%%%%%%%%%%%%%%%%%%%%%%%%%%%%%%%%%%%%%%%%%%%%%%%%%%%%%%%%%%%
% In the PDF output, floats should be placed
% under their own appendix, not before the title, nor after the
% title of the next appendix.
\onecolumn
\section{Light curves}
Table \ref{tab:LC} presents the optical observed $uBVgri$ magnitudes acquired with the Swope Telescope. Magnitudes are given in the Swope's natural system. 

\begin{table}[ht!]
    \centering
    \caption{Optical light curves of SN~2024ggi.}
    \label{tab:LC}
    \begin{tabular}{lccccccc}
    \hline
        MJD & Phase & u & B & V & g & r & i \\
        \hline
60984.4 & 573.6 & ... & ... & ... & ... & $19.96 \pm 0.10$ & ... \\
61087.4 & 676.6 & ... & $21.79 \pm 0.07$ & $21.70 \pm 0.07$ & $21.63 \pm 0.05$ & $20.84 \pm 0.04$ & $21.60 \pm 0.10$ \\
61097.2 & 686.4 & $22.19 \pm 0.14$ & $21.72 \pm 0.07$ & $21.67 \pm 0.08$ & $21.75 \pm 0.06$ & $20.93 \pm 0.04$ & $21.81 \pm 0.13$ \\
61108.1 & 697.3 & $22.21 \pm 0.23$ & $21.77 \pm 0.09$ & $21.84 \pm 0.12$ & $21.82 \pm 0.08$ & $20.88 \pm 0.04$ & $21.78 \pm 0.14$ \\
61115.2 & 704.4 & $22.34 \pm 0.15$ & $21.89 \pm 0.07$ & $21.80 \pm 0.09$ & $21.78 \pm 0.06$ & $20.91 \pm 0.05$ & $21.69 \pm 0.10$ \\
61122.2 & 711.4 & $22.91 \pm 0.18$ & $21.95 \pm 0.08$ & $21.90 \pm 0.10$ & $21.69 \pm 0.05$ & $20.98 \pm 0.05$ & ... \\
61129.1 & 718.3 & ... & $21.72 \pm 0.15$ & $22.07 \pm 0.21$ & $21.65 \pm 0.12$ & $21.07 \pm 0.08$ & $22.20 \pm 0.24$ \\
61138.1 & 727.3 & $23.23 \pm 0.21$ & $21.96 \pm 0.08$ & $21.81 \pm 0.09$ & $21.82 \pm 0.06$ & $20.98 \pm 0.05$ & $22.10 \pm 0.18$ \\
61152.2 & 741.4 & $23.03 \pm 0.23$ & $22.10 \pm 0.08$ & $21.97 \pm 0.10$ & $21.82 \pm 0.04$ & $21.00 \pm 0.04$ & $22.31 \pm 0.20$ \\
61229.0 & 818.2 & ... & $21.72 \pm 0.06$ & $22.12 \pm 0.12$ & $21.73 \pm 0.05$ & $21.07 \pm 0.04$ & $22.14 \pm 0.14$ \\
        \hline
    \end{tabular}
\end{table}

\end{appendix}
\end{document}